\documentclass[aip, showpacs, reprint, preprintnumbers,amsmath,amssymb,superscriptaddress,floatfix]{revtex4-2}
\usepackage{graphicx}      
\usepackage{bm}             	
\usepackage{times}			
\usepackage{tabularx}
\usepackage{color}
\usepackage{dcolumn}		
\usepackage{amsmath}
\usepackage{amssymb}
\usepackage{amsfonts}
\usepackage{times}
\usepackage{xcolor,colortbl}
\usepackage{amsmath}
\usepackage{multirow}
 \usepackage{ragged2e}
\usepackage[T1]{fontenc}
\usepackage[autostyle]{csquotes}
   \MakeAutoQuote{‘}{’}
\usepackage{csquotes} 
\usepackage[english]{babel}
\usepackage{ulem}
\definecolor{lime}{HTML}{A6CE39}
\usepackage{orcidlink}

\makeatletter
\newcommand\xleftrightarrow[2][]{%
  \ext@arrow 9999{\longleftrightarrowfill@}{#1}{#2}}
\newcommand\longleftrightarrowfill@{%
  \arrowfill@\leftarrow\relbar\rightarrow}
\makeatother
\setcitestyle{square}

\begin{document}

\title{Symmetry of the dissipation of surface acoustic waves by ferromagnetic resonance}  

\author{Florian Millo \orcidlink{0000-0002-8248-6635}}
\email{florian.millo@cnrs.fr}
\affiliation{Universit\'e Paris-Saclay, C2N, CNRS, 91120 Palaiseau, France}
\author{Rafael Lopes Seeger \orcidlink{0000-0001-9683-9553}}
\affiliation{SPEC, CEA, CNRS, Université Paris-Saclay, 91191 Gif-sur-Yvette, France}
\author{Claude Chappert \orcidlink{0000-0002-0129-8158}}
\affiliation{Universit\'e Paris-Saclay, C2N, CNRS, 91120 Palaiseau, France}
\author{Aur\'elie Solignac \orcidlink{0000-0002-7714-0939}}
\affiliation{SPEC, CEA, CNRS, Université Paris-Saclay, 91191 Gif-sur-Yvette, France}
\author{Thibaut Devolder \orcidlink{0000-0001-7998-0993}}
\email{thibaut.devolder@cnrs.fr}
\affiliation{Universit\'e Paris-Saclay, C2N, CNRS, 91120 Palaiseau, France}
\date{\today}                                           
\begin{abstract}
We study the symmetry of the coupling between surface acoustic waves and ferromagnetic resonance in a thin magnetic film of CoFeB deposited on top of a piezoelectric Z-cut LiNbO$_3$ substrate. We vary the orientation of the applied magnetic field with respect to the wavevector of the surface acoustic wave. Experiments indicate an unexpected 2-fold symmetry of the absorption of the SAW energy by the magnetic film. We discuss whether this symmetry can arise from the magnetoelastic torque of the longitudinal strain and the magnetic susceptibility of ferromagnetic resonance. We find that one origin of the 2-fold symmetry can be the weak in-plane uniaxial anisotropy present within the magnetic film. This phenomena adds to the previously identified other source of 2-fold symmetry but shall persist for ultrathin films when the dipolar interactions cease to contribute to the anisotropy of the slope of the spin wave dispersion relation.    
\end{abstract}


\maketitle 

Magnetoelasticity deals with the coupling of the mechanical strain and the magnetic degrees of freedom. In particular, it couples surface acoustic waves (SAW) and ferromagnetic resonance (FMR) as theoretically discussed in ref.~[\onlinecite{kittel_interaction_1958}].
Experimental demonstrations of this coupling were typically done on heterostructures comprising first thick [\onlinecite{bommel_excitation_1959}, \onlinecite{pomerantz_excitation_1961}], then thin [\onlinecite{weiler_elastically_2011, dreher_surface_2012, thevenard_surface-acoustic-wave-driven_2014,labanowski_power_2016}] magnetic films deposited on top of piezoelectric substrates. The piezoelectric materials used in literature are often treated as isotropic and the SAW-FMR coupling is generally found to arise at four evenly spaced field orientations [\onlinecite{weiler_elastically_2011}]. This 4-fold symmetry of SAW-FMR coupling is thought to result directly from the symmetry of the magnetoelastic torque of the longitudinal strain induced by the SAW on the magnetic film [\onlinecite{dreher_surface_2012, thevenard_surface-acoustic-wave-driven_2014,labanowski_power_2016}].
Huang et al. [\onlinecite{huang_large_2024}] have identified another contribution to the symmetry of the coupling of SAW and spin waves: A coupling with 2-fold symmetry can arise because the dispersion relations of the spin waves have this same symmetry. For instance, backward volume spin waves may be resonant with the SAWs while magnetostatic surface spin waves may not. Besides, a weak uniaxial anisotropy of the magnetic layers was recently shown to also influence the symmetry of the SAW-FMR coupling ~[\onlinecite{chen_widen-dynamic-range_2024}].

In this paper, we study the symmetry of the SAW-FMR coupling in SAW devices comprised of magnetostrictive Co$_{40}\textrm{Fe}_{40}\textrm{B}_{20}$ (CoFeB) deposited on top of a Z-cut LiNbO$_3$ substrate. We find an unexpected 2-fold symmetry of SAW-FMR coupling, that is maximal at field orientations that do not arise from arguments based on the anisotropy of spin wave dispersion relations. We develop a model that takes into account the weak uniaxial anisotropy of the magnet and reproduces the main experimental trends. \\ 
\indent The paper is organized as follows. Sect.~\ref{secexperiments} describes our samples and experimental setup, as well as the obtained symmetry of the SAW-FMR coupling. In Sect.~\ref{symmetriesofmagnetoacousticC}, we discuss the physical effects at play in SAW-FMR coupling and their potential symmetries. Sect.~\ref{theory} describes a model based on Smit \& Beljers formalism [\onlinecite{smit_ferromagnetic_1955}] and discuss the expected impact of magnetic anisotropy on the SAW-FMR coupling. In Sect.~\ref{bestfittheoryexp}, we compare the measurements with the outcomes of the model. We then summarize the main conclusions.
\section{Experimental study of the dissipation of surface acoustic waves by coupling to the magnetization}\label{secexperiments} 
\begin{figure*}[!t]
    \centering
    \includegraphics[scale=0.2]{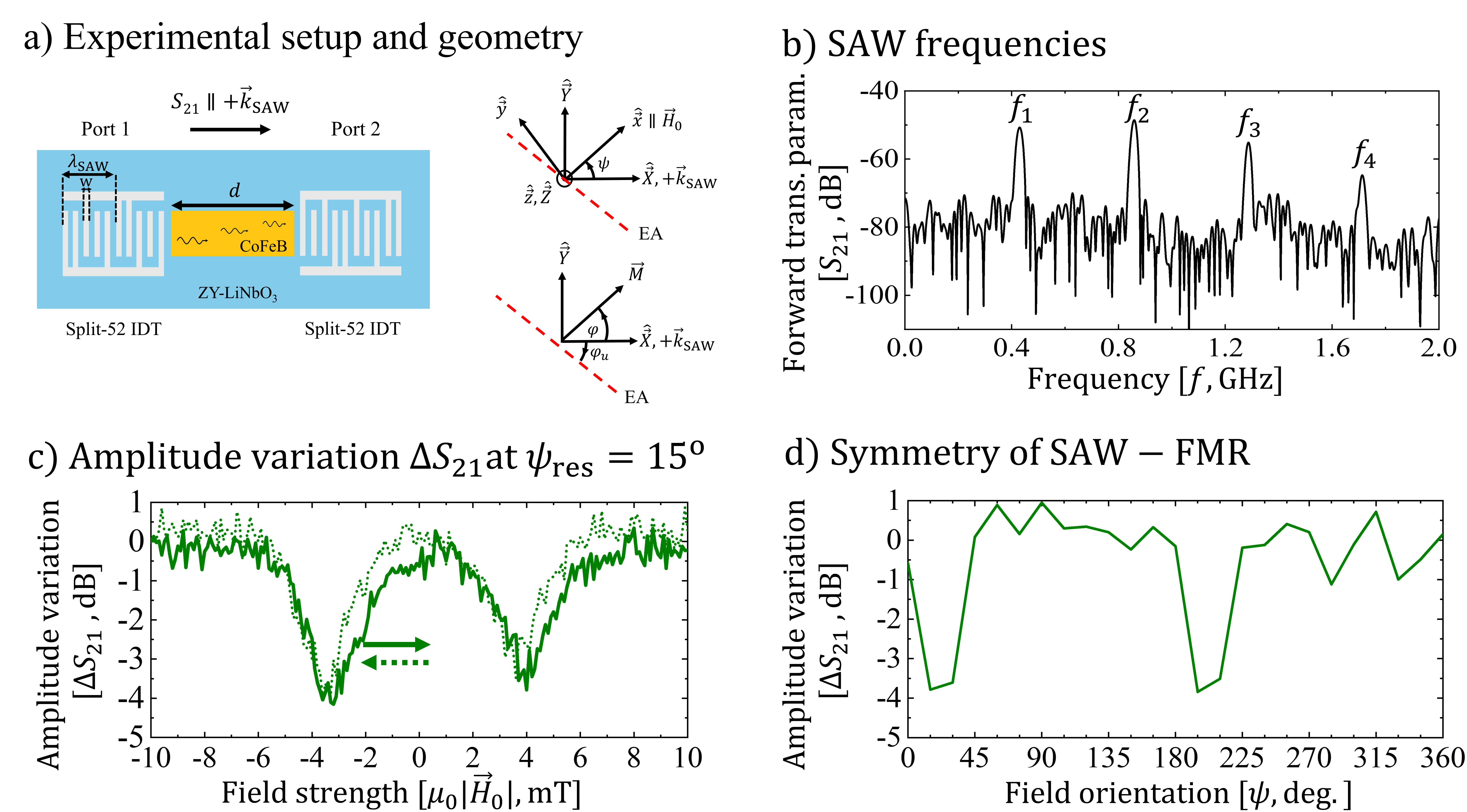}
    \caption{ a) Experimental configuration and definitions. The SAWs are excited using a split-52 IDT using the first port of a VNA and the transmission is collected at the second port. The IDTs periodicity is $\lambda_{\textrm{SAW}}=9\;\mu$m and the finger width is $w=0.9\;\mu$m. A CoFeB rectangle of length $d=800\;\mu$m is placed between the IDTs. \{$xyz$\} is the reference frame of the in-plane dc field $\vec{H}_0 \parallel \hat{\vec x}$, and \{$XYZ$\} is the reference frame of the LiNbO$_3$ crystal. EA stands for easy axis. b) Example of time-gated transmission signal $S_{21}$ defining the SAW frequencies $f_n=n f_1$ GHz for  $n=$ 1, 2, 3, 4 and $f_1=0.43$ GHz. c) Field dependence of $\Delta S_{21}$ for a field oriented at $\psi=15^\circ$ and at $f_4=1.72$ GHz. The arrows sketch the field sweeping direction. 
    d) Measured maximum (negative) value of $\Delta S_{21}$ at $f_4$ and $\mu_0|\vec{H}_0|=4$ mT versus the orientation of the applied field. Note the 2-fold symmetry. }
    \label{ExpSetupandAmpVarPanel}
\end{figure*}

We study the SAW-FMR coupling in the following material system: Z-cut LiNbO$_3$ / Ta(6 nm, buffer) / CoFeB($t=34$ nm) / Ru(0.4 nm) / Ta(3 nm, cap). The SAWs in this material and this specific crystalline orientation are very close to Rayleigh SAWs [\onlinecite{lopes_seeger_symmetry_2024}]. They are generated using 70 nm-thick aluminium interdigitated transducers (IDTs) of fundamental frequency $f_1=430$ MHz. A split-52 design [\onlinecite{schulein_fourier_2015}] is used 
to efficiently excite the odd and even harmonic SAWs at frequencies $f_n= n f_1$, $n \in \mathbb N^*$ [see Fig.~\ref{ExpSetupandAmpVarPanel}(a)]. The SAWs are launched in the Y crystalline direction of LiNbO$_3$. We pattern the CoFeB films in a rectangular structure between the two IDTs. 
Its properties include a magnetization of $\mu_0M_s=1.71$ T and in-plane uniaxial anisotropy of $\mu_0H_u=1.5$ mT [\onlinecite{mouhoub_exchange_2023}] oriented with an angle $\varphi_u$. The magnetoelastic coefficients of CoFeB are assumed to be $B_{1} \approx B_{2} = -7.6\times10^{6}$ J/m$^3$ [\onlinecite{masciocchi_strain-controlled_2021}]. 

The field dependence of the SAW transmission parameter $S_{21}$ is measured with a Vector Network Analyzer (VNA). 
The $S_{21}$'s are systematically post-processed using a time-gating procedure [\onlinecite{devolder_measuring_2021}] with gate start $0.23\; \mu$s and gate stop $0.3\;\mu$s. This removes the electromagnetic feed-through and improves the signal-to-noise ratio, revealing clearly the SAW harmonics [Fig.~\ref{ExpSetupandAmpVarPanel}(b)]. The frequencies $f_n$ of the SAW harmonics are consistent with a SAW phase velocity $V_{\textrm{SAW}}=$ 3870 m/s, in-line with the expected value for a Z-cut Y-propagating LiNbO$_3$ material [\onlinecite{kushibiki_chemical_2002}].

The measurements are done for applied in-plane dc fields $\mu_0|\vec{H}_{0}| \in [-10, 10]$ mT of variable orientations $\psi \in [0^\circ, 360^\circ]$. The information of SAW-FMR coupling is encoded in the variation $\Delta S_{21}$ [dB] of the IDT-to-IDT transmission parameter defined as:
\begin{align}
	\Delta S_{21}(H_0, \psi, f)  = 20 \log_{10} \left( \frac{|S_{21}(H_0, \psi, f)|}{|S_{21}(H_{\textrm{ref}}, \psi, f )|}\right)\label{varamplitude}
\end{align}
where the $\mu_0 H_{\textrm{ref}}=-10$ mT is a reference magnetic field chosen far from the SAW-FMR resonance fields. Fig.~\ref{ExpSetupandAmpVarPanel}(c) shows the hysteretic behaviour of the SAW-FMR coupling $\Delta S_{21}(H_0, f)$ for a field orientation at $\psi=15^\circ$ and at the frequency $f_4$. The coupling is maximal at $\mu_0|\vec{H}_0|\approx \pm 3.5-4$ mT, and leads to an extra loss of 4 dB [see Fig.~\ref{ExpSetupandAmpVarPanel}(c)].
The maximum of the extra loss depends on the field orientation $\psi$ with respect to the SAW wavevector direction [see Fig.~\ref{ExpSetupandAmpVarPanel}(d)]. In contrast to the commonly reported 4-fold symmetry, we find that $\Delta S_{21}(H_0, \psi, f)$ exhibits a clear 2-fold symmetry: the SAW energy is significantly absorbed only for two field orientations $\psi=15$ and 195$^\circ$. The next section inquires which physical effects can lead to this peculiar symmetry.

\section{Symmetries of the magneto-acoustic coupling} \label{symmetriesofmagnetoacousticC}

\begin{figure}[!t]
    \includegraphics[width=8.8cm]{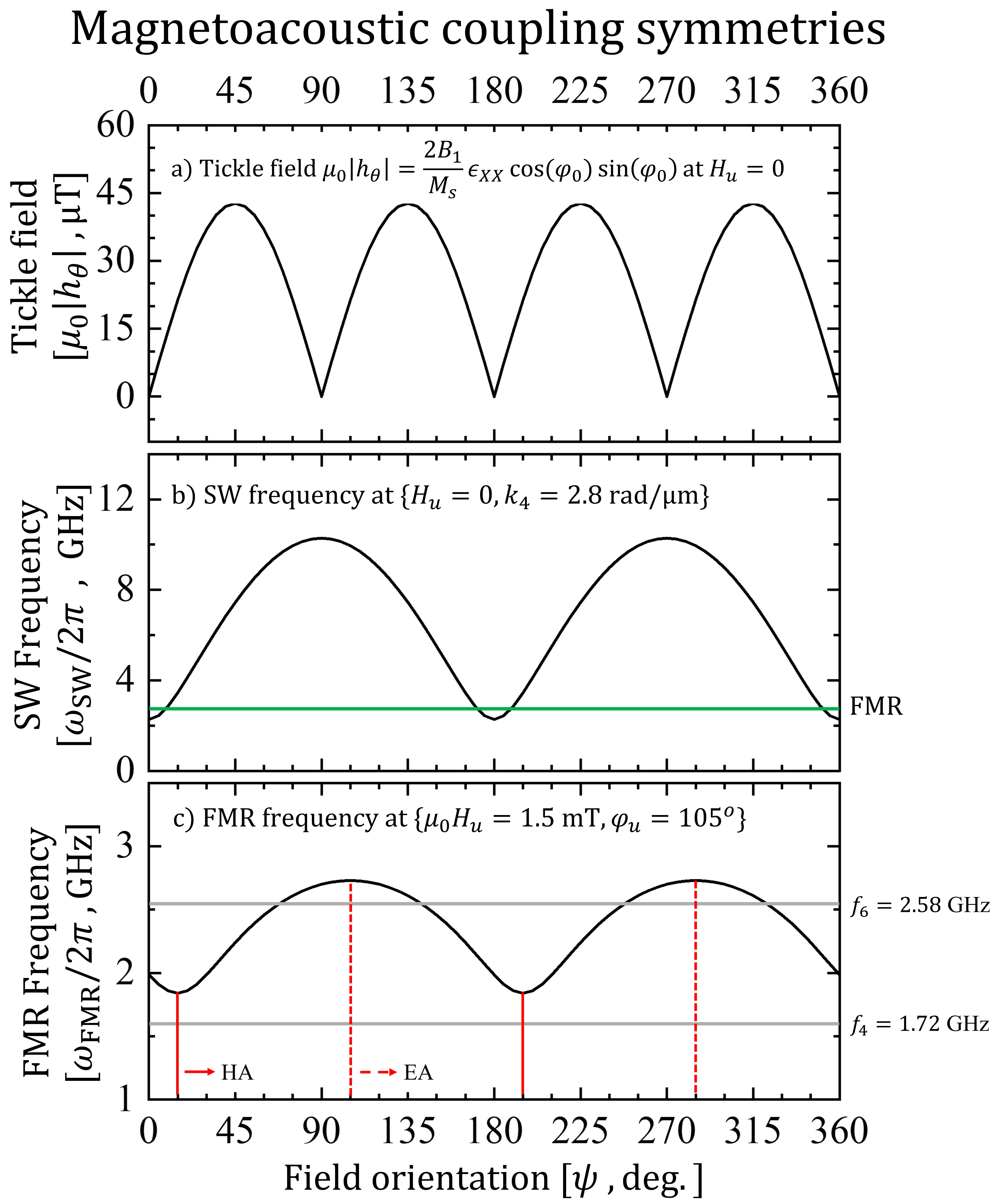}
    \caption{Illustration of the symmetries at play in 
    magnetoelastic coupling, evaluated for a field of $\mu_0|\vec{H}_0|=4$ mT of variable orientation, and for a CoFeB magnetic film of thickness 34 nm. a) Tickle field (torkance) for an isotropic magnetic film subjected to a sole longitudinal strain according to ref.~[\onlinecite{weiler_elastically_2011}]. b) Spin waves (SW) frequency [\onlinecite{kalinikos_theory_1986}] evaluated at SW wavevector $k_4 = 2.8$ rad/$\mu$m for an isotropic magnetic film. The green line represents the FMR $\omega_{\textrm{SW}}(k=0)$ value. c) FMR frequency evaluated for a uniaxial anisotropy field $\mu_0H_u=1.5$ mT oriented at $\varphi_u=105^\circ$. The vertical lines represent the easy and hard axes (EA, HA).}
    \label{couplingsymmetries}
\end{figure}

Past studies have identified different sources of the observed symmetry in the coupling of SAW to magnetization dynamics. These sources are conveniently revealed by the angular dependence of the three metrics plotted in Fig.~\ref{couplingsymmetries}: \\
\indent (i) The first metric is the magnetoelastic field arising from the longitudinal strain $\epsilon_{xx}$, also called tickle field or torkance. It occurs in an isotropic magnetic film at uniform resonance [Fig.~\ref{couplingsymmetries}(a)]. In this magnetically isotropic case, the equilibrium magnetization orientation matches with that of the field ($\varphi_0 = \psi$) and the magnetoelastic field has a perfect 4-fold symmetry that can be calculated following Weiler et al. [\onlinecite{weiler_elastically_2011}].  \\
\indent (ii) The second relevant metric is the spin wave resonance frequency of isotropic magnetic film for spin waves (SWs) of finite wavevectors [Fig.~\ref{couplingsymmetries}(b)], that can be calculated following Kalinikos and Slavin [\onlinecite{kalinikos_theory_1986}]. The SW resonance frequency exhibits a 2-fold symmetry with maxima at $\psi=\pm \pi/2$. Since the magnetic susceptibility peaks when in resonant conditions, the SAW-SW coupling shall reflect this 2-fold symmetry, as discussed in ref.  [\onlinecite{huang_large_2024}]. For didactic purposes, we shall not include this effect in our forthcoming model, since we aim at discussing the specific role of uniaxial anisotropy. \\
\indent (iii) The last relevant metric is the frequency of the uniform resonance (FMR-mode) in magnetic films with uniaxial anisotropy [Fig.~\ref{couplingsymmetries}(c)], that can be calculated according to Smit \& Beljers [\onlinecite{smit_ferromagnetic_1955}]. This FMR has a 2-fold symmetry versus field orientation. This symmetry should also appear in the angular dependence of the SAW-FMR coupling, as we shall model in the next section.
\section{Model of the dissipation of the SAW energy by FMR}\label{theory}

\subsection{Formalism} 
We use the Smit \& Beljers formalism [\onlinecite{smit_ferromagnetic_1955}] to describe the ferromagnetic resonance. The total energy $\mathcal{E}$ [J/m$^3$] for the configuration given in Fig.~\ref{ExpSetupandAmpVarPanel}(a) is [\onlinecite{hernandez-minguez_large_2020}],
\begin{align}
	\mathcal{E} &= -\mu_{0} M_{s} |\vec{H}_{0}| \left[\cos(\varphi-\psi)\right] +\frac{\mu_{0}M^{2}_{s}}{2} \cos^{2}(\theta)  \nonumber \\
	&- H_{u} \frac{\mu_0 M_{s}}{2}\sin^{2}(\theta)\cos^{2}(\varphi-\varphi_{u}), \label{sphericaletot}
\end{align}
where the first term is the Zeeman energy, the second term is the demagnetization energy, and the third term is the in-plane uniaxial anisotropy. Numerical energy minimization yields the ground state of the magnetization ($\vec{M}$), described in the usual spherical coordinates by $\theta_0$ (inclination from +$\hat{\vec{z}}$ axis) and $\varphi_0$ (revolutions in the $xy$ plane). The small fluctuations $\delta \theta \ll 1$ and $\delta \varphi \ll 1$ about the ground state are described by,
\begin{align}
   \theta &= \theta_0 + \delta \theta \nonumber \\
   \varphi &= \varphi_0 + \delta \varphi.
\end{align}

Under the application of rf fields $\{h_\theta, h_\varphi\}$, the small perturbations of the magnetization can then be expressed as,
\begin{align}
    	\frac{i \omega}{\gamma}\left(\begin{array}{cc}
		\delta \theta \\
        \delta \varphi
	\end{array}\right) = \overleftrightarrow{\chi}     	\left(\begin{array}{cc}
		h_\theta \\
        h_\varphi
	\end{array}\right). \label{magnandrffields}
\end{align}
In Eq.~\ref{magnandrffields}, $\overleftrightarrow{\chi}$ is the magnetization susceptibility defined as [\onlinecite{hernandez-minguez_large_2020}],
\begin{align}
   \overleftrightarrow{\chi} &= \frac{(\gamma/M_{s})^2}{(1+\alpha^2)\omega^2-\omega^2_{\textrm{FMR}}+  \frac{i \alpha \omega \gamma}{M_s} (\mathcal{E}_{\theta \theta}+\mathcal{E}_{\varphi \varphi}) } \nonumber \\
	&\times
	\left(
	\begin{array}{cc}
		\mathcal{E}_{\varphi \varphi} - \frac{ i \alpha \omega M_{s}}{ \gamma }	 & \frac{i \omega  M_{s}}{\gamma} \\
		-\frac{i\omega  M_{s}}{\gamma} &  \mathcal{E}_{\theta \theta} -  \frac{i\alpha\omega  M_{s}}{\gamma} \\
	\end{array}
	\right)\label{susceptibility}
\end{align}
where $\gamma>0$ is the gyromagnetic ratio, $\omega=2\pi f_{\textrm{SAW}}$ is the SAW frequency, $\omega_{\textrm{FMR}}$ is the FMR frequency and the subscripts of $\mathcal{E}$ mean their second partial derivatives with respect to $\theta$ and $\varphi$. For in-plane magnetized film $\theta_0=\pi/2$ and $\varphi_{0}$ are found by numerically minimizing Eq.~\ref{sphericaletot}. The FMR frequency $\omega_{\textrm{FMR}}\big|_{\varphi=\varphi_0}$ is defined as,
\begin{align}
	\omega_{\textrm{FMR}} &=\gamma \mu_0 \sqrt{|\vec{H}_{0}| \cos(\varphi_0-\psi) + H_{u}  \cos[2(\varphi_0-\varphi_u)]} \nonumber\\ &\times \sqrt{|\vec{H}_{0}| \cos(\varphi_0-\psi) + M_{s}+   H_{u} \cos^2(\varphi_0-\varphi_u) } \label{almostkittel}.
\end{align}

In this study we consider only the longitudinal strain of the SAW, hence the magnetoelastic energy (mel) is [\onlinecite{lopes_seeger_symmetry_2024}],
\begin{align}
    \mathcal{E}_{\textrm{mel}} = B_{1} \epsilon_{xx}m^2_{x}. \label{melenergy}
\end{align}
The SAW-FMR coupling is studied in the reference frame \{$xyz$\} of the magnetic field $\vec{H}_0$. Hence a proper transformation from \{$XYZ$\} to \{$xyz$\} reference frame is performed in a similar way as in 
refs.~[\onlinecite{weiler_elastically_2011}, \onlinecite{lopes_seeger_symmetry_2024}]. We then deduce from Eq.~\ref{melenergy}, the in-plane tickle field $h_\theta$ generated by SAW as,
\begin{align}
    h_{\theta}&= \frac{2 B_1}{\mu_0 M_s} \epsilon_{XX} \cos{(\varphi_0)} \sin{(\psi)},
\end{align}
where $\epsilon_{XX}=0.76\times10^{-5}$ is a typical value for a Z-cut LiNbO$_3$ material [\onlinecite{lopes_seeger_symmetry_2024}]. Since we consider in-plane magnetized films, the $h_{\varphi}$ component of the magnetoelastic field vanishes at first order \footnote{Note that we disregard the shear component $\epsilon_{xz}$ of the strain and therefore cannot account for the non-reciprocity of the coupling [\onlinecite{dreher_surface_2012, hernandez-minguez_large_2020, xu_nonreciprocal_2020}].}.
To quantify the SAW-FMR coupling we use the time-averaged power transmitted to the magnetic film (W/m$^2$, power per unit surface where SAWs are present) [\onlinecite{lopes_seeger_symmetry_2024}],
\begin{align}
\Delta P=-\frac{\mu_0 \omega t}{2} \textrm{Im} \left( h_{\theta}^{\dagger} \cdot\overleftrightarrow{\chi} \cdot h_\theta \right)\Big|_{\varphi=\varphi_0}, \label{powertransmitted}
\end{align}
where the dagger symbol means transpose-conjugate.
\subsection{Effect of the strength of uniaxial magnetic anisotropy on SAW-FMR coupling} \label{secvarHu}
\begin{figure}[!t]
    \includegraphics[width=8.9cm]{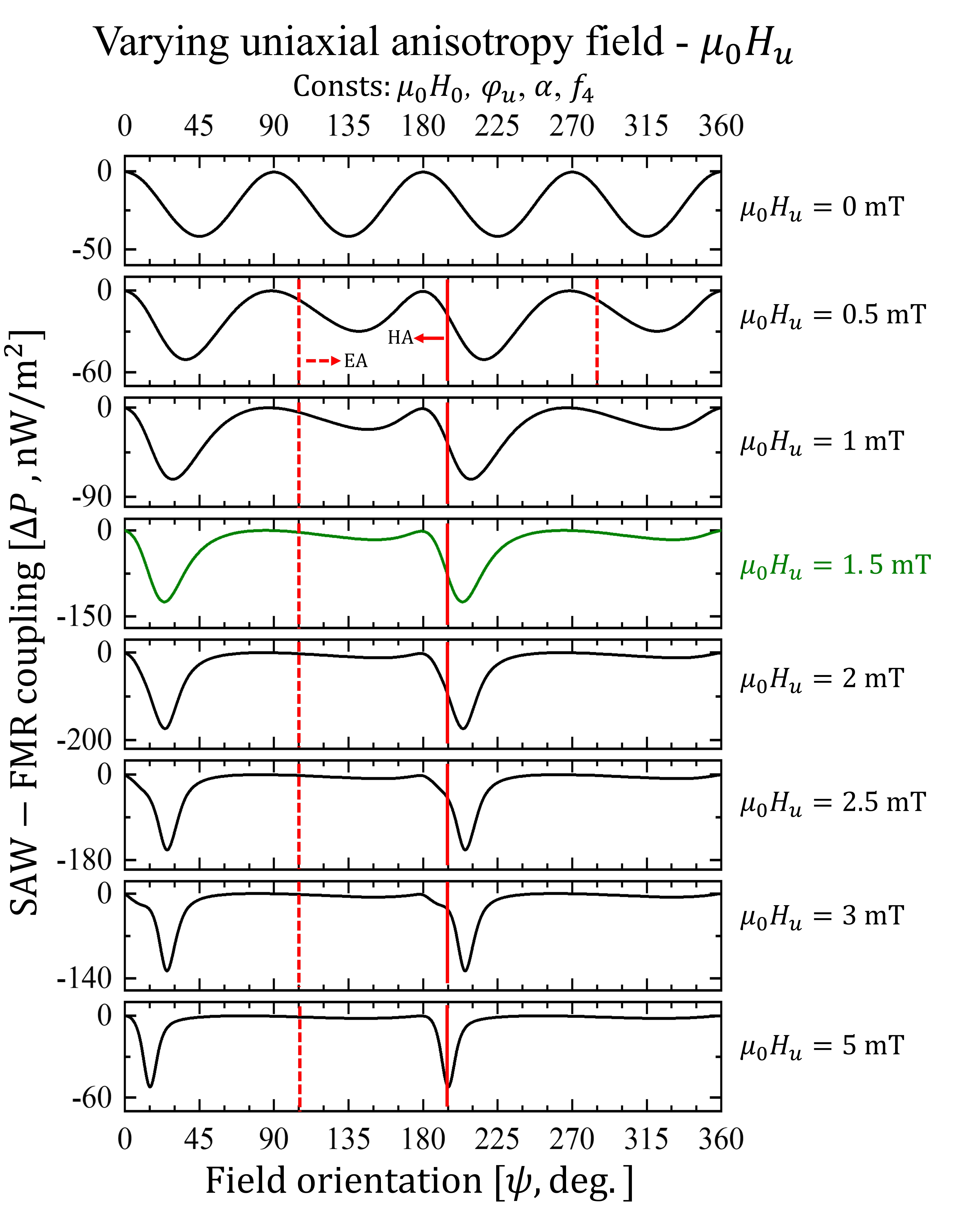}
    \caption{Angular dependence of SAW-FMR coupling when varying the uniaxial anisotropy field $\mu_0H_u$ and keeping constants $\mu_0|\vec{H}_0|=4$ mT, uniaxial anisotropy angle $\varphi_u=105^\circ$, $\alpha=0.01$ and SAW frequency $f_4=1.72$ GHz. The vertical lines represent the easy and hard axes (EA, HA). The green curve is the best fit of the measurements.}
    \label{varHu}
\end{figure}
We first investigate the SAW-FMR coupling when there is an in-plane uniaxial anisotropy of variable strength [Fig.~\ref{varHu}]. The specific role of $H_u$ can be revealed by keeping constant the direction $\varphi_u=105^\circ$ of the easy axis and the SAW frequency. 
The isotropic case ($H_u=0)$ [see the top of Fig.~\ref{varHu}] leads to the well-known 4-fold symmetry where $\forall \psi,\, \Delta P (\psi)= \Delta P (\psi+ n\pi/2)$ with $n\in \mathbb{Z}$, with the maximum coupling occurring at $\psi_{\textrm{res}}=\pi/4+n\pi/2$ [\onlinecite{weiler_elastically_2011}]. 
However by incrementing $H_u$ the 4-fold symmetry is progressively lost while a 2-fold symmetry is always maintained, i.e., $\forall \psi,\, \Delta P (\psi)= \Delta P (\psi+ n\pi)$ still holds. For small anisotropy, the maximum coupling occurs in between the previous $\pi/4+n\pi/2$ orientations and the hard axis of the anisotropy. The coupling is maximal for anisotropy fields of typically 2 mT. At anisotropy fields much larger than the applied field (last panel in Fig.~\ref{varHu}) the maximum coupling finally tends to align with the hard axis. Since the orientation of the field leading to maximum coupling can change, it is worth studying the impact of the orientation $\varphi_u$ of the easy axis.
 
\begin{figure}[!t]
    \includegraphics[width=8.75cm]{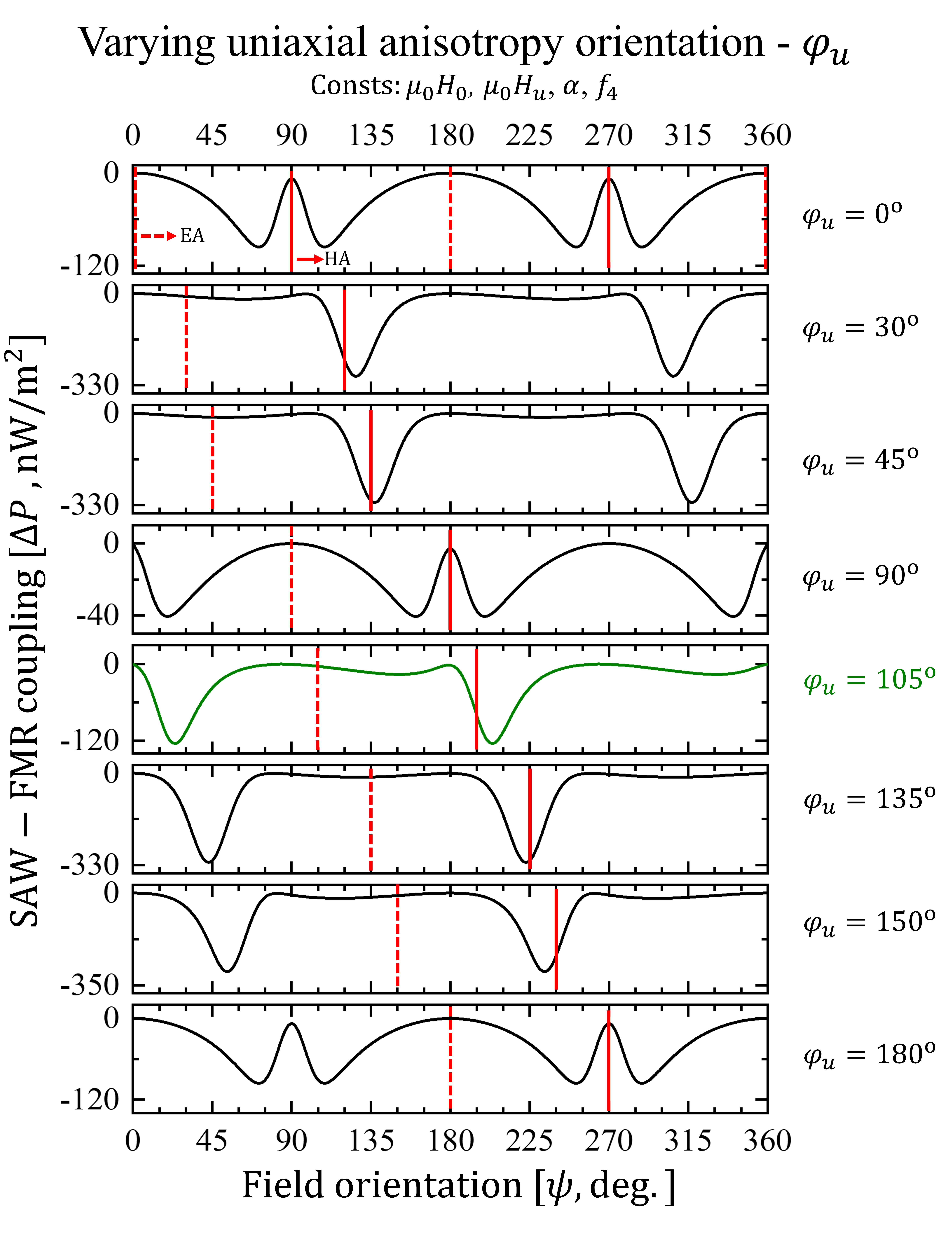}
    \caption{Angular dependence of SAW-FMR coupling when varying the uniaxial anisotropy angle $\varphi_u$ and keeping constants $\mu_0|\vec{H}_0|=4$ mT, uniaxial anisotropy field $\mu_0H_u=1.5$ mT, $\alpha=0.01$ and SAW frequency $f_4=1.72$ GHz. The vertical lines represent the easy and hard axes (EA, HA). The green curve is the best fit of the measurements.}
    \label{varphiu}
\end{figure} 
\subsection{Effect of the orientation of uniaxial magnetic anisotropy on SAW-FMR coupling}
We now investigate the SAW-FMR coupling when rotating the easy axis. Fig.~\ref{varphiu} shows that as soon as there is some uniaxial anisotropy, whatever its orientation, the SAW-FMR coupling looses its 4-fold symmetry and only the 2-fold symmetry remains. 
It is worth noticing that when either of the principal axes of the uniaxial magnetic anisotropy is along $x$ (i.e. for $\varphi_u=n \pi /2 $ for $n\in\mathbf{Z}$), there are 4 orientations leading to maximum SAW-FMR coupling.  Two of the maxima progressively disappear when departing from these orientations. 
The parameters chosen in Fig.~\ref{varphiu} include $H_0 > H_u$ such that $\omega_{\textrm{FMR}}(\psi) > \omega_4$ for $\forall \varphi_u$ of the easy axis. This results in maximum coupling that is always near the hard axis. We will come back to this point in the discussion.
Before proceeding with this discussion, we also investigate the frequency dependence of the angular-resolved SAW-FMR coupling.

\subsection{Effect of the frequency on the SAW-FMR coupling}
\begin{figure}[!t]
    \includegraphics[width=8.7cm]{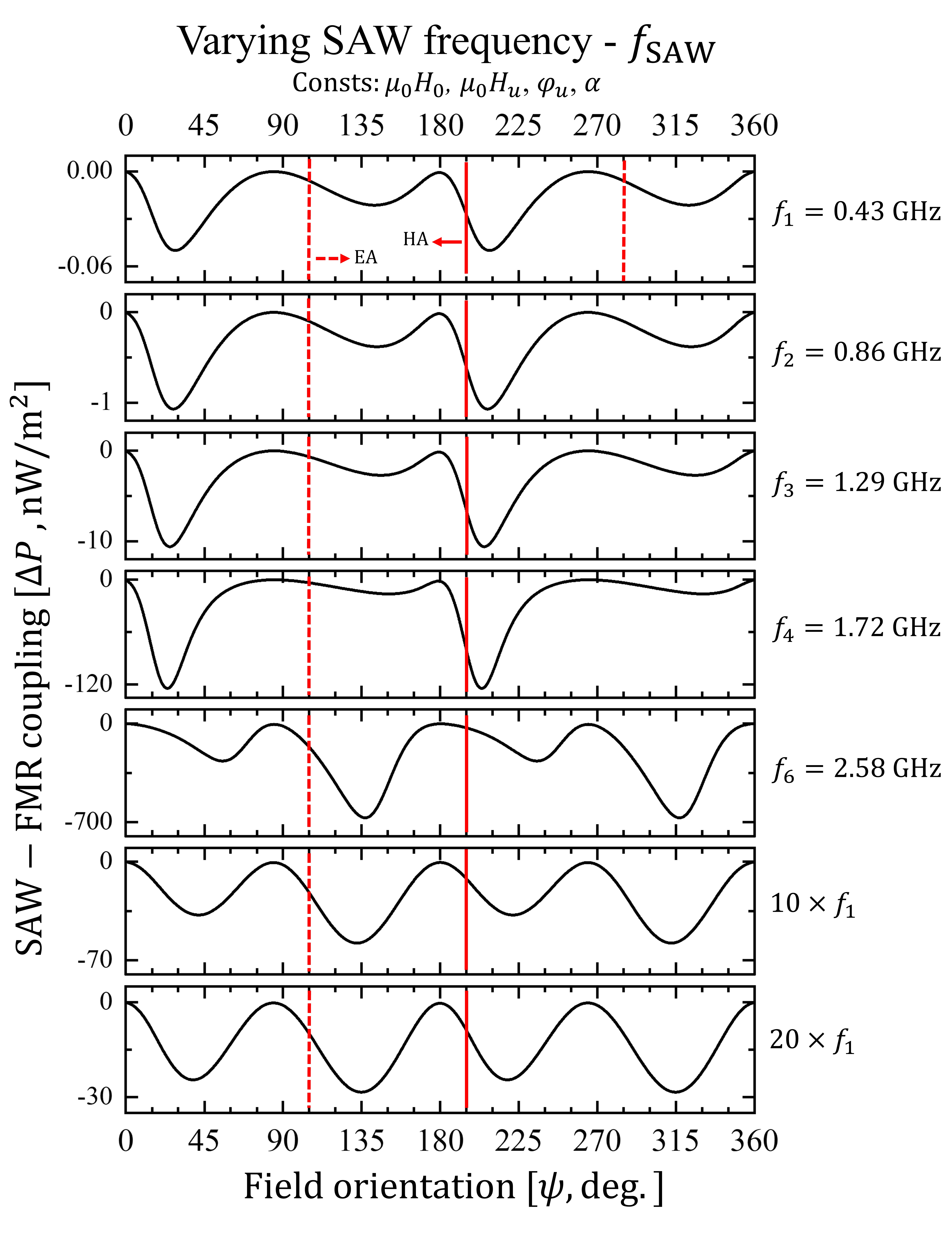}
    \caption{Angular dependence of SAW-FMR coupling when varying the SAW frequency $f_{\textrm{SAW}}$, and keeping constants $\mu_0|\vec{H}_0|=4$ mT, uniaxial anisotropy field $\mu_0H_u=1.5$ mT, the uniaxial anisotropy angle $\varphi_u=105^\circ$ and $\alpha=0.01$. The vertical lines represent the easy and hard axes (EA, HA).}
    \label{varf}
\end{figure}
We will finally discuss the symmetry of SAW-FMR coupling when varying the SAW frequency $f_{\textrm{SAW}}$ at which the coupling is evaluated [Fig.~\ref{varf}].
At the lowest SAW frequency $f_{\textrm{SAW}}=f_1=0.43$ GHz, the SAW is very far from being resonant with the FMR and the coupling is consequently very low [see the top panel of Fig.~\ref{varf}]. The maximum coupling is achieved for applied fields oriented close to the hard axis. The coupling is much stronger at intermediate frequencies when there exists field orientations for which the SAW can be resonant with the FMR. Since the FMR frequency depends on the field orientation, the coupling peaks at field orientations that depends strongly on the applied frequency. Finally at very large frequencies, the SAW can no longer be resonant with the FMR such that the magnitude of the coupling decreases and the 4-fold symmetry is almost recovered. The maxima of the coupling tend to correspond to the orientations of maximum torkance of the isotropic system [see Fig.~\ref{couplingsymmetries}(a)], i.e. they occur at orientations near $\pi/4$, $3\pi/4$, $5\pi/4$ and $7\pi/4$ [see the bottom panel of Fig.~\ref{varf}] with little dependence on the orientation of the easy axis (not shown).    

\begin{figure*}[!t]
    \centering
    \includegraphics[width=17cm]{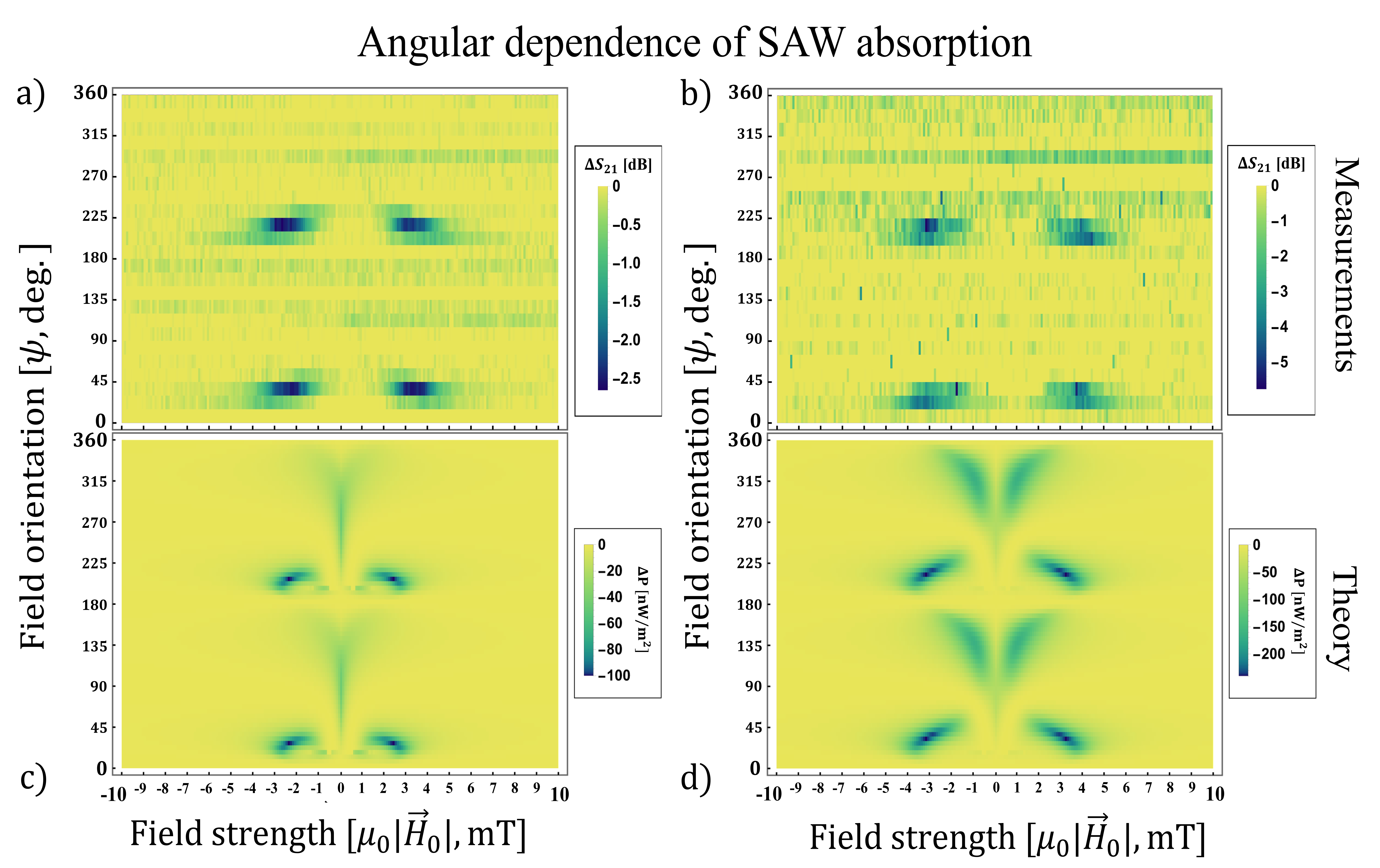}
     \caption{Field orientation and field strength dependence of Y-propagating SAW absorption in the heterostructure comprised of Z-cut LiNbO$_3$ / CoFeB(34 nm). Measurements ($\Delta S_{21}$ [dB]) conducted at a) $f_3=1.29$ GHz and at b) $f_4=1.72$ GHz. Theory ($\Delta P$ [nW/m$^{2}$]) implemented at c) $f_3=1.29$ GHz and at d) $f_4=1.72$ GHz. Fitted parameters are $\mu_0H_u=1.5$ mT, $\varphi_u=105^\circ$, and $\alpha=0.01$.}
    \label{2DmapsAngularDependency}
\end{figure*}
\subsection{Practical Consequences of the model}
In summary, our model predicts that a 4-fold symmetry of the SAW-FMR coupling is only possible for magnetically isotropic material. As soon as some magnetic uniaxial anisotropy is present in the sample and whatever its orientation, the symmetry is reduced to 2-fold.

The magnitude of the SAW-FMR coupling can be understood from the angular dependence of the torkance [Fig.~\ref{couplingsymmetries}(a)] and from the detuning between the SAW and the FMR-mode frequencies [Fig.~\ref{couplingsymmetries}(c)]. The magnitude of the coupling is typically low in large detuning conditions, i.e., when the applied frequency is much below, or much above the FMR for all field orientations. In this case, the (weak) maxima of the coupling tend to coincide with the orientations of the maximum torkance of the isotropic system [Fig.~\ref{couplingsymmetries}(a)]. The coupling is maximal for magnetization orientations $\varphi_0$ at $\pi/4$, $3\pi/4$, $5\pi/4$ and $7\pi/4$. For $H_0 \gg H_u$ the magnetization is almost aligned with the applied field such that, this maximum coupling tends to correspond to the same field orientations, $\psi$ at $\pi/4$, $3\pi/4$, $5\pi/4$ and $7\pi/4$. 

The SAW-FMR coupling is strong when the applied frequency can hit the FMR, except when this requires a magnetization orientation that happen to coincide with a vanishing torkance, i.e., when $\varphi_0= n\,\pi/2$. In a correlated manner, the strongest maxima of the coupling occur when the SAW frequency matches the FMR for a field orientation that leads to one of the high torkance orientations ($\varphi_0= \pi/2 + n \pi/4$). 

This can be used to formulate a rule of thumb to ensure large coupling. A very good coupling can be expected when $H_0 = H_u$, $\varphi_u=0$ and when the SAW frequency is set in-between the easy axis frequency $\gamma_0 \sqrt{(H_0 + H_u)(H_0 + H_u+M_s)} $ and the hard axis frequency $\gamma_0 \sqrt{(H_0 - H_u)(H_0 - H_u+M_s)} $. Indeed in this case, the matching of the FMR resonance with $f_{SAW}$ is obtained for a field orientation that is in between the easy and the hard axis, which leads to a magnetization orientation that is also in between the easy and the hard axis. We then have approximately $\psi\approx \varphi_0 \approx \pi/4$ which coincides with the maximum torkance, thereby providing strong SAW-FMR coupling.\\
\indent As a last comment, let us examine the situation of large applied fields, i.e. $H_0 > H_u$. In this case, rotating the applied field along a full turn $\psi \in [0 \rightarrow 2 \pi]$ triggers a full continuous turn of the magnetization $\varphi_0 \in [0 \rightarrow 2 \pi]$. As a result, the system crosses the zero torkance magnetization orientations $\varphi_0= n\,\pi/2$ four times, such that regardless of the matching or non-matching of the resonances, the coupling must vanish 4 times during this full rotation of the applied field. This happens in all panels of Figs.~\ref{varphiu} and \ref{varf}, as well as in all panels of Fig.~\ref{varamplitude} except the very bottom one. Any experimental deviation from this "four zeros" rule is indicative that other components of the strain tensor are magnetoelastically active. \\
\indent Although we are aware that our model neglects the anisotropy of the SW dispersion relation [Fig.~\ref{couplingsymmetries}(b)] as well as the contributions from the shear strain and from the magneto-rotation coupling [\onlinecite{lopes_seeger_symmetry_2024}], we will compare the predictions of SAW-FMR coupling to experimental measurements to clarify what our model captures and what it misses.
\section{Comparison between measurements and theory} \label{bestfittheoryexp}
In this section we compare the measurements with the model developed in Sect.~\ref{theory}. Fig.~\ref{2DmapsAngularDependency} reports the dependence of the SAW-FMR coupling versus field amplitude and orientation, i.e., maps of the experimental $\Delta S_{21}(H_0, \psi)$ and the modeled $\Delta P(H_0, \psi)$ for the SAW frequencies $f_3$ and $f_4$. A damping of $\alpha=0.01$ was assumed to account for the linewidths seen in the measurements. The model has successes and limitations. \\
\indent The model clearly reproduces the 2-fold symmetry of the coupling. The high coupling zones are visible in all panels of Fig.~\ref{2DmapsAngularDependency} as dark tilted segments emerging from uniformly colored regions with no coupling. These coupling maxima are predicted at the correct field orientations ($\psi_{\textrm{res}}=15^\circ\sim 30^\circ+n\,180^\circ$) and for the correct field strengths (e. g. $\mu_0|\vec{H}_{0,\;\textrm{res}}|=3\sim5$ mT). The increase of the coupling with increasing SAW frequency is also clearly reproduced. For instance both the  predicted $\Delta P$ and the measured $|\Delta S_{21}|$ double when passing from $f_3$ to $f_4$ \footnote{The agreement between experiments and modeling is also correct at $f_1=0.43$ GHz and $f_2=0.86$ GHz, for which the fields leading to maximum coupling are $\mu_0|\vec{H}_{0,\;\textrm{res}}|=2\sim3$ mT (not shown)}.\\
\indent Despite these successes, the model wrongly predicts a faint but finite coupling at low fields for large intervals of orientations near $\psi=135^\circ\;\textrm{and}\;315^\circ$. This is seen as $\vee$-shaped halos in Fig.~\ref{2DmapsAngularDependency}(d) and their narrower counterparts in Fig.~\ref{2DmapsAngularDependency}(c). These halos are absent from the experimental data. This limitation of the model can be understood from Fig.~\ref{couplingsymmetries}(b). In the regions where $\psi \approx \varphi_0 \approx 0~\textrm{or~}\pi$, the spin waves have a backward volume character with almost vanishing group velocity such that their frequency is very close to FMR and thus quite well described by the Eq.~\ref{almostkittel} used in the model. The predictions are thus correct when they concern these field orientations and spin waves of backward volume character. In contrast, the spin waves have a magnetostatic surface wave character (hence with large group velocity) in the regions where $\psi \approx \varphi_0 \approx \pi/2$, such that the spin wave frequency are very different from the FMR frequency used in the model. The real detuning is thus substantially underestimated by Eq.~\ref{almostkittel} for these magnetization orientations. Accounting for the real detuning would substantially suppress the coupling for these orientations and likely account for the experimentally observed values. 
\section{Summary and conclusions} 
We have investigated the coupling between surface acoustic waves propagating in the Y crystalline direction of a Z-cut LiNbO$_3$ substrate, and the magnetization of the CoFeB thin film. Our experiments show that the dissipation of the SAW energy by its coupling to the magnetic film depends on the strength and on the orientation of the applied magnetic field, with a clear 2-fold symmetry. This symmetry cannot be accounted for by the sole angular variation of the magnetoelastic torque of the longitudinal strain.

Using a model based on the magnetic susceptibility of the ferromagnetic resonance, we show that this 2-fold symmetry can arise from the combination of the magnetoelastic torque of the longitudinal strain and the presence of some in-plane uniaxial anisotropy within the magnetic material. The model is meant to evidence the role of uniaxial anisotropy and therefore does neither take into account the contributions of shear strains, nor that of the lattice rotations, which are of both of secondary importance for thin in-plane magnetized systems like ours. 
Our model is nominally restricted to the sole uniform ferromagnetic resonance, but is applicable whenever the involved spin waves have frequencies close to the uniform resonance. The model will thus be applicable in the case in ultrathin films for magnetostatic spin waves, but when in thick films, it is restricted to the coupling of SAWs with spin waves of wavevectors close to the magnetostatic backward volume configuration. We have calculated how the strength of the uniaxial anisotropy, its orientation and the frequency of the used SAW frequency matter for the symmetry of the coupling. The model reproduces the experimental findings except in the regions where the involved spin waves have frequencies far from the uniform resonance.  Our findings can be used to formulate rules of thumb to ensure large SAW to spin wave coupling in the case of magnetic material exhibiting some uniaxial anisotropy.
\section*{Acknowledments}
The authors acknowledge the French National Research Agency (ANR) under contract N$^{o}$ ANR-20-CE24-0025 (MAXSAW). This work was supported by a public grant overseen by the French National Research Agency (ANR) as part of the “Investissements d'Avenir” program (Labex NanoSaclay, reference: ANR-10-LABX-0035, project SPICY). This work was done within the C2N micro nanotechnologies platforms and partly supported by the RENATECH network and the General Council of Essonne.%

\end{document}